\RequirePackage{fix-cm}
\documentclass[smallextended]{svjour3}       
\smartqed  
\usepackage{graphicx}
\usepackage[utf8]{inputenc}
\usepackage{amsmath}
\usepackage[nodisplayskipstretch]{setspace}
\usepackage{authblk}
\usepackage{hyperref}
\usepackage[none]{hyphenat}
\usepackage{qcircuit}
\usepackage{float}
\usepackage{braket}

\begin{document}

\title{Efficient schemes for  the quantum teleportation of a sub-class of tripartite entangled states
}


\author{Dintomon Joy         \and
        Sabir M 
}


\institute{Dintomon Joy \and Sabir M \at
              Department of Physics, Cochin University of Science and Technology, Kochi - 682 022, India \\
              \email{dintomonjoy@cusat.ac.in}           
               }

\date{Received: date / Accepted: date}

\maketitle

\begin{abstract}
\sloppy In this paper we propose two schemes for teleportation of a sub-class of tripartite states, the first one
 with the four-qubit cluster state and the second one with two Bell pairs as entanglement channels.
 A four-qubit joint measurement in the first case and two Bell measurements in the second are performed by the sender.
 Appropriate unitary operations on the qubits at the receiver's end along with an ancilla qubit result in the perfect
 teleportation of the tripartite state. Analysis of the quantum circuits employed in these schemes reveal that in our
 technique the desired quantum tasks are achieved with lesser quantum cost, gate count and classical communication bits
 compared with other similar schemes.

\end{abstract}

\section{Introduction}
\label{intro}

\sloppy Quantum telportation is an important and interesting task which has recently attracted much attention because of its
promising applications in the field of secure quantum communication and quantum computing. In the teleportation process
an unknown quantum state is recreated at a distant location by exploiting shared entanglement and classical communication
between the sender and the receiver. In the original scheme of Bennett \textit{et al.} \cite{bennett}, verified first
experimentally by Bouwmeester \textit{et al.} \cite{Bowmeester}, it was shown that an unknown single-particle quantum state can be teleported using a maximally entangled Bell pair as a quantum  channel. Following these developments, research interest in this area was focused on other schemes for single-qubit teleportation as well the possibilities of teleporting
multipartite states.

Various schemes for teleporting bipartite entangled states \cite{marinatto,cao&song,Cola,Chuang} as well as  arbitrary
bipartite states \cite{Rigolin,yeo} have been suggested. Studies on teleportation of three-particle entangled states
gained momentum when Liu and Guo \cite{liu02} first showed that probabilistic teleportation of a general three-particle
GHZ state is possible with three non-maximally entangled Bell pairs. It has also been reported that perfect teleportation of
an arbitrary three-particle states is possible via genuinely entangled six-qubit state\cite{yin} and also via three sets
of W-class states\cite{yi}. The question whether it is possible to achieve perfect teleportation of sub-classes of tripartite states with the involvement of lesser entanglement resources has also been discussed. Pradhan\textit{et al.}\cite{pradhan}, Nie \textit{et al.}\cite{Nie} and Li \textit{et al.}\cite{YXLM16} demonstrated teleportation of certain sub-classes of three particle states (with two unknown coefficients) using genuinely entangled four-qubit cluster states as entanglement channel. Liu and Zhou \cite{zhong} worked out a scheme for teleportation of a specific form of tripartite state (with four unknown coefficients) using five qubit cluster state. Chaudhary and Dhara\cite{BA16} showed that the tripartite states of the form similar to the one used in \cite{zhong} can be teleported using three sets of four-qubit cluster states. Later, Wei \textit{et.al.}\cite{WZY} showed that the same task in \cite{BA16} can be achieved using only one four-qubit entangled state as entanglement channel. In a recent work, Cai and Jiang\cite{CJ} showed a much improved scheme for the tasks given in \cite{BA16}\cite{WZY}.

In this work, we present two  efficient schemes and explicit quantum circuits to realize the teleportation of the special forms of three-qubit states with four unknown coefficients. Of these the first method uses  the four-qubit cluster state and the second one uses two Bell pairs as entanglement channels. In  our methods,  a four-qubit joint measurement in the first case and two Bell measurements in the second case are done by the sender and specified unitary operations
  on the  qubits at the receiver's end along with an ancilla qubit results in the perfect teleportation of the
  tripartite state. This is based on the observation that performing a Hadamard operation and a measurement on
  any one of the qubits of given tripartite state by the sender does not destroy the information about the unknown
  coefficients. By a detailed analysis of the quantum circuits employed in these schemes we show that in our protocols the desired quantum tasks are achieved with lesser requirements on quantum cost, gate count and classical communication bits
compared with other similar schemes.

This paper is organized as follows. In section:\ref{sec:1}, after a brief discussion of the tripartite states, we propose a teleportation scheme using a four-qubit cluster state as entanglement channel. In section:\ref{sec:2}, we use two Bell pairs for teleportation of the same tripartite state. In section:\ref{sec:3}, we compare our schemes with other similar schemes and in the final section we present our conclusions.

\section{Scheme for teleportation of special type of tripartite state using four-qubit cluster state}
\label{sec:1}
Recently, Zhang \textit{et.al}\cite{zhang} and Yang \textit{et.al}\cite{yang} pointed out the possibility of using special
forms of GHZ-like channels for the controlled teleportation of an unknown qubit. Generalizing these results, Pathak and
Banerjee \cite{Anirban} showed that there exists 12 maximally entangled GHZ-like states having a general structure,
\begin  {equation}
\label{eqn: 1}
\left|\Gamma_{123}\right\rangle=\frac{1}{\sqrt{2}}\{\left|\psi_{i}\right\rangle_{12}\left|0\right\rangle_3 + \left|\psi_{j}\right\rangle_{12}\left|1\right\rangle_3\}
\end{equation}
where, i, j = \{0, 1, 2, 3\}, i $\neq$ j and $\left|\psi_{i}\right\rangle$, $\left|\psi_{j}\right\rangle$ represents Bell states; $\left|\psi_{0}\right\rangle = \left|\phi^{+}\right\rangle=
\frac{1}{\sqrt{2}}\{\left|00\right\rangle + \left|11\right\rangle\}$,
$\left|\psi_{1}\right\rangle =\left|\psi^{+}\right\rangle= \frac{1}{\sqrt{2}}\{\left|01\right\rangle + \left|10\right\rangle\}$,
 $\left|\psi_{2}\right\rangle = \left|\phi^{-}\right\rangle =
\frac{1}{\sqrt{2}}\{\left|00\right\rangle - \left|11\right\rangle\}$, and $\left|\psi_{3}\right\rangle =\left|\psi^{-}\right\rangle= \frac{1}{\sqrt{2}}\{\left|01\right\rangle - \left|10\right\rangle\}$,  which can be
used as quantum channels for the controlled teleportation of n-qubit non-maximally entangled states of generalized
Bell-type. If we replace the Bell states $\left|\psi_{i}\right\rangle$ and $\left|\psi_{j}\right\rangle$ in
Eq.(\ref{eqn: 1}) with non-maximally entangled states
 $ \left|\psi_0'\right\rangle=a_0 \left|00\right\rangle + b_0 \left|11\right\rangle$, $ \left|\psi_1'\right\rangle=a_1 \left|01\right\rangle + b_1 \left|10\right\rangle$, $ \left|\psi_2'\right\rangle= a_2 \left|00\right\rangle + b_2 \left|11\right\rangle$ and $ \left|\psi_3'\right\rangle= a_3 \left|01\right\rangle + b_3 \left|10\right\rangle$ 
with normalized coefficients $({|a_0|}^2 + {|b_0|}^2) = ({|a_1|}^2 + {|b_1|}^2) = ({|a_2|}^2 + {|b_2|}^2) = ({|a_3|}^2 + {|b_3|}^2)=1$,
we get the following four unique non-maximally entangled arbitrary tripartite states with four different terms
\begin{eqnarray}
\label{four}
\left|\Psi_A\right\rangle_{123} = a\left|010\right\rangle+b\left|100\right\rangle+c\left|011\right\rangle+d\left|101\right\rangle_{123}\\\nonumber
\left|\Psi_B\right\rangle_{123} = a\left|010\right\rangle+b\left|100\right\rangle+c\left|001\right\rangle+d\left|111\right\rangle_{123}\\\nonumber
\left|\Psi_C\right\rangle_{123} = a\left|000\right\rangle+b\left|110\right\rangle+c\left|001\right\rangle+d\left|111\right\rangle_{123}\\ \nonumber
\left|\Psi_D\right\rangle_{123} = a\left|000\right\rangle+b\left|110\right\rangle+c\left|011\right\rangle+d\left|101\right\rangle_{123}\\ \nonumber
\end{eqnarray}
where, $a, b, c$ and $d$ are obtained by redefining the coefficients (with ${|a|}^2 + {|b|}^2 + {|c|}^2 + {|d|}^2=1$). These three-particle states are related to each other by unitary transformations $ \sigma_{x_{2}} \left|\Psi_A\right\rangle_{123} = \left|\Psi_C\right\rangle_{123} $, $ \sigma_{x_{2}} \left|\Psi_B\right\rangle_{123} = \left|\Psi_D\right\rangle_{123} $, CNOT$(3,2)\left|\Psi_A\right\rangle_{123} = \left|\Psi_B\right\rangle_{123}$ and CNOT$(3,2)\left|\Psi_C\right\rangle_{123} = \left|\Psi_D\right\rangle_{123}$, where, $\{I, \sigma_{x}, \sigma_{y}, \sigma_{z}\}$ represents Pauli matrices. The teleportation of the tripartite state $\left|\Psi_C\right\rangle_{123}$ has been discussed in \cite{zhong}, \cite{BA16}, \cite{WZY} and \cite{CJ}. In this section, we investigate the possibility of teleporting all these special class of arbitrary tripartite states given in equation:\ref{four}, using four-qubit cluster state as entanglement channel. The four-qubit cluster state (Equation:~\ref{eqn:1}) introduced by Briegel and Raussendorf\cite{briegel} is known to have maximal connectedness and large amount of entanglement.
\begin{equation}
 \label{eqn:1}
 \left|\Psi\right\rangle_{4567}= 1/2\{\left|0000\right\rangle+ \left|0110\right\rangle+\left|1001\right\rangle -\left|1111\right\rangle\}_{4567}\\
\end{equation}

In our protocol, the sender Alice and the receiver Bob  share two particles each from the four-qubit
cluster state prior to the teleportation process. Alice is in possession of particles $(1,
2, 3, 4, 5)$ and Bob has particles $6, 7$ and an ancilla qubit 8. Suppose, Alice wants to teleport the state $\left|\Psi_{B}\right\rangle_{123}$ to Bob. To achieve this, initially, Alice applies a Hadamard operation on particle $1$ in her possession. This operation puts the state of $\left|\Psi_{B}\right\rangle_{123}$ in a superposition state as shown below.
\begin{equation}
 \label{eqn:2}
\left|\Psi_B\right\rangle_{123} = \left|0\right\rangle_1 (a\left|10\right\rangle+b\left|00\right\rangle+c\left|01\right\rangle +d\left|11\right\rangle)_{23}+ \left|1\right\rangle_1 (a\left|10\right\rangle-b\left|00\right\rangle+
c\left|01\right\rangle-d\left|11\right\rangle)_{23}
\end{equation}
 \begin{figure}
\[
\Qcircuit @C=.55em @R=.8em {
\lstick{\ket{1}} & \qw  & \qw  & \qw  &\gate{H} & \meter   & \cw & \cw & \cw & \cw & \cw & \cw & \cw & \cw\cw  & \cw & \cw \cw & \cw  &\cw &\control{7} \cw &\cw \\
\lstick{\ket{2}} & \qw  & \qw & \qw  & \qw & \qw  & \qw & \ctrl{3}  &\qw  &\qw & \gate{H}   &\meter & \cw & \control \cw & \cw & \cw  & \cw & \cw  & \cw\cwx &\cw\\
\lstick{\ket{3}}  & \qw  & \qw  & \qw  & \qw & \qw  & \qw & \qw  & \qw & \ctrl{1} & \gate{H} &\meter &\cw & \cw \cwx  & \cw  &\control \cw &\cw & \cw & \cw \cwx  & \cw\\
\lstick{\ket{4}} &\gate{H} &\qw &\qw  &\ctrl{3} &\qw &\ctrl{1} &\qw &\qw &\targ    & \qw  &\meter & \cw  & \cw \cwx & \control \cw  & \cw \cwx  & \cw  & \cw &\cw \cwx & \cw\\
\lstick{\ket{5}} &\gate{H} &\ctrl{1} & \qw & \qw  & \qw  & \gate{Z} &\targ &\qw  &\qw &\qw \gategroup{2}{7}{5}{11}{.5em}{--} & \meter &\control \cw &\cw \cwx &\cw \cwx &\cw \cwx &\cw &\cw &\cw \cwx &\cw\\
\lstick{\ket{6}}  & \qw &\targ  & \qw  & \qw &\ctrl{1}  & \qw  & \qw   & \qw & \qw  & \qw  & \qw &\gate{X}\cwx &\gate{Z}\cwx &\qw \cwx &\qw \cwx &\qw &\ctrl{2} &\qw \cwx &\qw \\
\lstick{\ket{7}}  & \qw  & \qw  & \qw & \targ   & \gate{Z} \gategroup{4}{2}{7}{6}{.5em}{--}  & \qw & \qw  & \qw  & \qw &\qw  &\qw &\qw &\qw &\gate{X}\cwx &\gate{Z}\cwx &\ctrl{1} & \qw & \qw \cwx & \qw & & & & \ket{\Psi_B}_{867}\\
\lstick{\ket{8}}  & \qw  & \qw  & \qw & \qw  & \qw  & \qw & \qw  & \qw  & \qw &\qw & \qw & \qw & \qw & \qw & \qw &\targ \gategroup{7}{17}{8}{17}{.6em}{.} & \targ &\gate{Z}\cwx \gategroup{6}{18}{8}{20}{.6em}{\}} &\qw\\
 }\]
\caption{ The proposed quantum circuit for the teleportation of tripartite state using Four-qubit cluster state. The first dashed box shows the preparation of entanglement channel. The second dashed box represents four-particle joint measurement.  The dotted CNOT gate between qubits 7 and 8 is not utilized in the procedure for recovering the quantum states $\ket{\Psi_A}$ and $\ket{\Psi_C}$. }
\label{fig:2}
 \end{figure}
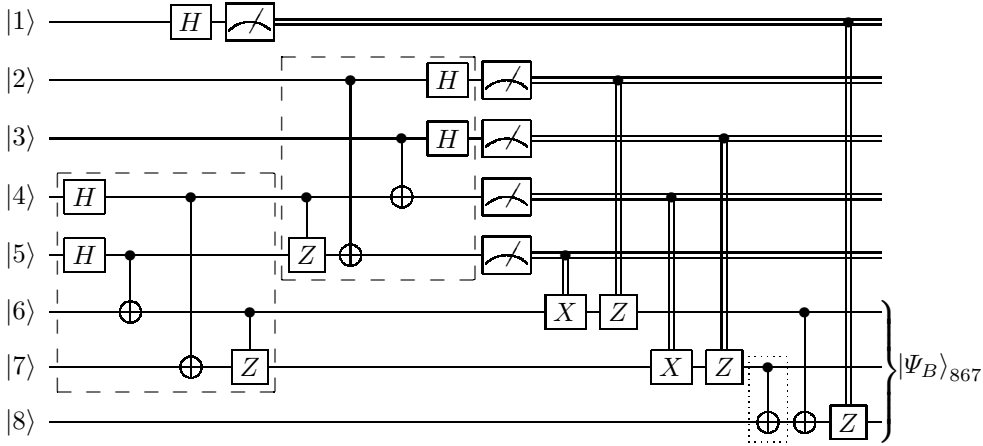
Next, Alice performs a single particle measurement on qubit $1$ and sends her result to Bob via classical communication. Later, using this information, Bob will perform a $Z$ operation on his ancilla qubit to recover the teleported state. After Alice's measurement on qubit $1$, the state of the system $(2,3)$ collapses to one of the two possible arbitrary two particle states as given in the equation (\ref{eqn:2}). Let us first, consider the case where the single particle measurement result of qubit $1$ is $\left|1\right\rangle_1$. Then the combined state of the system is given by $\left|\Psi\right\rangle_{tot} = \left|{1}\right\rangle_1 \otimes \left|\Phi\right\rangle_{23} \otimes \left|\Psi\right\rangle_{4567}$, where $\left|\Phi\right\rangle_{23}= a\left|10\right\rangle-b\left|00\right\rangle+c\left|01\right\rangle
-d\left|11\right\rangle)_{23}$. Then, Alice performs a four-qubit joint measurement on her particles $(2,3,4,5)$
in the measurement basis (as in \cite{XXJ16}) given below:
\begin{eqnarray*}
\left|\phi\right\rangle_{1,2}    &= 1/2 \{\left|0000\right\rangle + \left|0110\right\rangle \pm \left|1001\right\rangle \mp \left|1111\right\rangle\}_{1245}\\
\left|\phi\right\rangle_{3,4}    &= 1/2 \{\left|0000\right\rangle - \left|0110\right\rangle \pm \left|1001\right\rangle \pm \left|1111\right\rangle\}_{1245}\\
\left|\phi\right\rangle_{5,6}    &= 1/2 \{\left|0001\right\rangle + \left|0111\right\rangle \pm \left|1000\right\rangle \mp \left|1110\right\rangle\}_{1245}\\
\left|\phi\right\rangle_{7,8}    &= 1/2 \{\left|0001\right\rangle - \left|0111\right\rangle \pm \left|1000\right\rangle \pm \left|1110\right\rangle\}_{1245}\\
\left|\phi\right\rangle_{9,10}   &= 1/2 \{\left|0010\right\rangle + \left|0100\right\rangle \pm \left|1011\right\rangle \mp \left|1101\right\rangle\}_{1245}\\
\left|\phi\right\rangle_{11,12}  &= 1/2 \{\left|0010\right\rangle - \left|0100\right\rangle \pm \left|1011\right\rangle \pm \left|1101\right\rangle\}_{1245}\\
\left|\phi\right\rangle_{13,14}  &= 1/2 \{-\left|0011\right\rangle - \left|0101\right\rangle \mp \left|1010\right\rangle \pm \left|1100\right\rangle\}_{1245}\\
\left|\phi\right\rangle_{15,16}  &= 1/2 \{-\left|0011\right\rangle + \left|0101\right\rangle \mp \left|1010\right\rangle \mp \left|1100\right\rangle\}_{1245}
\end{eqnarray*}
\begin{table}
\caption{Outcome of Alice's Joint measurement and resulting state of qubits $(6,7)$}
\label{tab:1}
\begin{tabular}{l l l }
\hline\noalign{\smallskip}
JM   & State of Qubits $(6,7)$ &  State of Qubits $(6,7)$ \\
 Result & when qubit $1$ is $\left|0\right\rangle_1$  & when qubit $1$ is $\left|1\right\rangle_1$  \\
$(2,3,4,5)$&  & \\
\noalign{\smallskip}\hline\noalign{\smallskip}
$\left|\phi_{1}\right\rangle$ &$+a\left|10\right\rangle+b\left|00\right\rangle+c\left|01\right\rangle+d\left|11\right\rangle$&$+a\left|10\right\rangle-b\left|00\right\rangle+c\left|01\right\rangle-d\left|11\right\rangle$\\
$\left|\phi_{2}\right\rangle$ &$-a\left|10\right\rangle+b\left|00\right\rangle+c\left|01\right\rangle-d\left|11\right\rangle$&$-a\left|10\right\rangle-b\left|00\right\rangle+c\left|01\right\rangle+d\left|11\right\rangle$ \\

$\left|\phi_{3}\right\rangle$ &$+a\left|10\right\rangle+b\left|00\right\rangle-c\left|01\right\rangle-d\left|11\right\rangle$&$+a\left|10\right\rangle-b\left|00\right\rangle-c\left|01\right\rangle+d\left|11\right\rangle$ \\
$\left|\phi_{4}\right\rangle$ &$-a\left|10\right\rangle+b\left|00\right\rangle-c\left|01\right\rangle+d\left|11\right\rangle$&$-a\left|10\right\rangle-b\left|00\right\rangle-c\left|01\right\rangle-d\left|11\right\rangle$\\

$\left|\phi_{5}\right\rangle$ &$+a\left|00\right\rangle+b\left|10\right\rangle-c\left|11\right\rangle-d\left|01\right\rangle$&$+a\left|00\right\rangle-b\left|10\right\rangle-c\left|11\right\rangle+d\left|01\right\rangle$\\
$\left|\phi_{6}\right\rangle$ &$-a\left|00\right\rangle+b\left|10\right\rangle-c\left|11\right\rangle+d\left|01\right\rangle$&$-a\left|00\right\rangle-b\left|10\right\rangle-c\left|11\right\rangle-d\left|01\right\rangle$\\

$\left|\phi_{7}\right\rangle$ &$+a\left|00\right\rangle+b\left|10\right\rangle+c\left|11\right\rangle+d\left|01\right\rangle$&$+a\left|00\right\rangle-b\left|10\right\rangle+c\left|11\right\rangle-d\left|01\right\rangle$\\
$\left|\phi_{8}\right\rangle$ &$-a\left|00\right\rangle+b\left|10\right\rangle+c\left|11\right\rangle-d\left|01\right\rangle$&$-a\left|00\right\rangle-b\left|10\right\rangle+c\left|11\right\rangle+d\left|01\right\rangle$\\

$\left|\phi_{9}\right\rangle$ &$-a\left|11\right\rangle+b\left|01\right\rangle+c\left|00\right\rangle-d\left|10\right\rangle$&$-a\left|11\right\rangle-b\left|01\right\rangle+c\left|00\right\rangle+d\left|10\right\rangle$\\
$\left|\phi_{10}\right\rangle$&$+a\left|11\right\rangle+b\left|01\right\rangle+c\left|00\right\rangle+d\left|10\right\rangle$&$+a\left|11\right\rangle-b\left|01\right\rangle+c\left|00\right\rangle-d\left|10\right\rangle$\\

$\left|\phi_{11}\right\rangle$&$-a\left|11\right\rangle+b\left|01\right\rangle-c\left|00\right\rangle+d\left|10\right\rangle$&$-a\left|11\right\rangle-b\left|01\right\rangle-c\left|00\right\rangle-d\left|10\right\rangle$\\
$\left|\phi_{12}\right\rangle$&$+a\left|11\right\rangle+b\left|01\right\rangle-c\left|00\right\rangle-d\left|10\right\rangle$&$+a\left|11\right\rangle-b\left|01\right\rangle-c\left|00\right\rangle+d\left|10\right\rangle$\\

$\left|\phi_{13}\right\rangle$&$-a\left|01\right\rangle+b\left|11\right\rangle-c\left|10\right\rangle+d\left|00\right\rangle$&$-a\left|01\right\rangle-b\left|11\right\rangle-c\left|10\right\rangle-d\left|00\right\rangle$\\
$\left|\phi_{14}\right\rangle$&$+a\left|01\right\rangle+b\left|11\right\rangle-c\left|10\right\rangle-d\left|00\right\rangle$&$+a\left|01\right\rangle-b\left|11\right\rangle-c\left|10\right\rangle+d\left|00\right\rangle$\\

$\left|\phi_{15}\right\rangle$&$-a\left|01\right\rangle+b\left|11\right\rangle+c\left|10\right\rangle-d\left|00\right\rangle$&$-a\left|01\right\rangle-b\left|11\right\rangle+c\left|10\right\rangle+d\left|00\right\rangle$  \\
$\left|\phi_{16}\right\rangle$&$+a\left|01\right\rangle+b\left|11\right\rangle+c\left|10\right\rangle+d\left|00\right\rangle$&$+a\left|01\right\rangle-b\left|11\right\rangle+c\left|10\right\rangle-d\left|00\right\rangle$\\
\noalign{\smallskip}\hline\noalign{\smallskip}
\end{tabular}
\end{table}

\begin{table}
\caption{Result of Alice's Joint measurement (JM) and corresponding unitary operations (UO) to be performed by Bob on qubits $(6,7)$}
\label{tab:2}
\begin{tabular}{l l }
\hline\noalign{\smallskip}
JM    & UO \\
Result of & on qubits$(6,7)$ \\
$(2,3,4,5)$& after JM \\
\noalign{\smallskip}\hline\noalign{\smallskip}
$\left|\phi_{1}\right\rangle$ &$ I_6 \otimes I_7$\\
$\left|\phi_{2}\right\rangle$ &$\sigma_{z_{6}} \otimes I_7 $\\

$\left|\phi_{3}\right\rangle$ &$I_6 \otimes \sigma_{z_{7}}$\\
$\left|\phi_{4}\right\rangle$ &$\sigma_{z_{6}} \otimes \sigma_{z_{7}} $\\

$\left|\phi_{5}\right\rangle$ &$\sigma_{x_{6}} \otimes \sigma_{z_{7}} $\\
$\left|\phi_{6}\right\rangle$ &$i\sigma_{y_{6}} \otimes \sigma_{z_{7}}$\\

$\left|\phi_{7}\right\rangle$ &$\sigma_{x_{6}} \otimes I_7 $\\
$\left|\phi_{8}\right\rangle$ &$i\sigma_{y_{6}} \otimes I_7 $\\

$\left|\phi_{9}\right\rangle$ &$\sigma_{z_{6}} \otimes \sigma_{x_{7}}$\\
$\left|\phi_{10}\right\rangle$&$I_6 \otimes \sigma_{x_{7}}$\\

$\left|\phi_{11}\right\rangle$&$\sigma_{z_{6}} \otimes i\sigma_{y_{7}}  $\\
$\left|\phi_{12}\right\rangle$&$I_6 \otimes i\sigma_{y_{7}}$\\

$\left|\phi_{13}\right\rangle$&$i\sigma_{y_{6}} \otimes i\sigma_{y_{7}}$\\
$\left|\phi_{14}\right\rangle$&$\sigma_{x_{6}} \otimes i\sigma_{y_{7}}$\\

$\left|\phi_{15}\right\rangle$&$i\sigma_{y_{6}} \otimes \sigma_{x_{7}}$\\
$\left|\phi_{16}\right\rangle$&$ \sigma_{x_{6}} \otimes \sigma_{x_{7}}$\\
\noalign{\smallskip}\hline
\end{tabular}
\end{table}

The measurement result of particles $(2,3,4,5)$ and the corresponding state of particles $(6,7)$ are given in table \ref{tab:1}. Depending on the measurement result communicated by Alice, Bob
performs unitary operations on qubits $(6,7)$ as given in table \ref{tab:2}.
Later, Bob entangles his ancilla bit prepared in the state $ \left|1\right\rangle_8$ with qubits 6 and 7 by
performing two CNOT operations. The qubits $6$ and $7$ act as control bits and qubit $8$ as target bit. Finally, a Z operation (based on classical information from Alice) as shown in figure \ref{fig:2} is applied on qubit $8$ to recover the
teleported state $\ket{\Psi_B}$. Similarly, for the other case where
the measurement result of qubit $1$ is $\left|0\right\rangle_1$, the state of qubits
(2,3) collapse to $(a\left|10\right\rangle+b\left|00\right\rangle+c\left|01\right\rangle+d\left|11\right\rangle)_{23}$.
By following the same procedure, recovery operations in table \ref{tab:2} and the same circuit given in figure:~\ref{fig:2}, Bob can recover the teleported state in the qubits $6,7,8$ as $\ket{\Psi_B}_{867}$.

In the case of teleporting the states $\ket{\Psi_A}, \ket{\Psi_C}$ and $\ket{\Psi_D}$, Alice and Bob adopts the same procedure given in figure:\ref{fig:2}. The same measurement basis and the recovery operations in table:\ref{tab:2} are used. But, to recover the different states, the receiver has to entangle ancilla qubit (prepared in $\ket{0}$ or $\ket{1}$ states) appropriately with the qubits 6 and 7 depending on the teleported state. For example, to recover the state $\ket{\Psi_D}$ the receiver entangles the ancilla qubit prepared in the state $\ket{0}_8$ with qubits $6$ and $7$. In the case of recovering $\ket{\Psi_A}$ and $\ket{\Psi_C}$ the ancilla qubits in the states $\ket{1}_8$ and $\ket{0}_8$ are used respectively. But, here the ancilla qubit is entangled only with qubit $6$ using a CNOT operation. The CNOT operation between 7 and 8 which is shown in dotted box in the figure:\ref{fig:2}, is not be applied and the teleported state can be recovered as $\ket{\Psi_B}_{867}$.

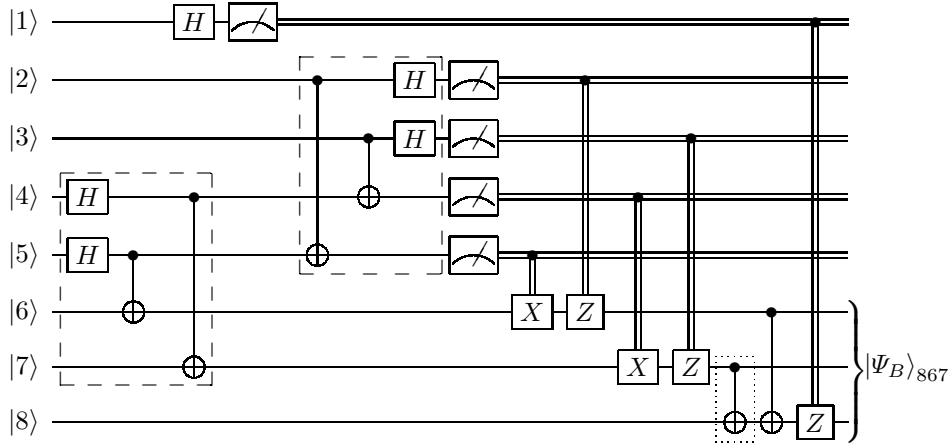
\begin{figure}
\[
\Qcircuit @C=.55em @R=.8em {
\lstick{\ket{1}} & \qw  & \qw  & \qw  &\gate{H} & \meter   & \cw & \cw & \cw & \cw & \cw & \cw & \cw & \cw\cw  & \cw & \cw \cw & \cw  &\cw &\control{7} \cw &\cw \\
\lstick{\ket{2}} & \qw  & \qw & \qw  & \qw & \qw  & \qw & \ctrl{3}  &\qw  &\qw & \gate{H}   &\meter & \cw & \control \cw & \cw & \cw  & \cw & \cw  & \cw\cwx &\cw\\
\lstick{\ket{3}}  & \qw  & \qw  & \qw  & \qw & \qw  & \qw & \qw  & \qw & \ctrl{1} & \gate{H} &\meter &\cw & \cw \cwx  & \cw  &\control \cw &\cw & \cw & \cw \cwx  & \cw\\
\lstick{\ket{4}} &\gate{H} &\qw &\qw  &\ctrl{3} &\qw &\qw &\qw &\qw &\targ    & \qw  &\meter & \cw  & \cw \cwx & \control \cw  & \cw \cwx  & \cw  & \cw &\cw \cwx & \cw\\
\lstick{\ket{5}} &\gate{H} &\ctrl{1} & \qw & \qw  & \qw  & \qw &\targ &\qw  &\qw &\qw \gategroup{2}{8}{5}{11}{.5em}{--} & \meter &\control \cw &\cw \cwx &\cw \cwx &\cw \cwx &\cw &\cw &\cw \cwx &\cw\\
\lstick{\ket{6}}  & \qw &\targ  & \qw  & \qw &\qw  & \qw  & \qw   & \qw & \qw  & \qw  & \qw &\gate{X}\cwx &\gate{Z}\cwx &\qw \cwx &\qw \cwx &\qw &\ctrl{2} &\qw \cwx &\qw \\
\lstick{\ket{7}}  & \qw  & \qw  & \qw & \targ   & \qw \gategroup{4}{2}{7}{5}{.5em}{--}  & \qw & \qw  & \qw  & \qw &\qw  &\qw &\qw &\qw &\gate{X}\cwx &\gate{Z}\cwx &\ctrl{1} & \qw & \qw \cwx & \qw & & & & \ket{\Psi_B}_{867}\\
\lstick{\ket{8}}  & \qw  & \qw  & \qw & \qw  & \qw  & \qw & \qw  & \qw  & \qw &\qw & \qw & \qw & \qw & \qw & \qw &\targ \gategroup{7}{17}{8}{17}{.6em}{.} & \targ &\gate{Z}\cwx \gategroup{6}{18}{8}{20}{.6em}{\}} &\qw\\
 }
\]
\caption{  Proposed quantum circuit for teleportation of tripartite state using two Bell pairs}
\label{fig:3}
 \end{figure}

\section{Scheme for teleportation of tripartite states using two Bell pairs as entanglement channel}
\label{sec:2}

In this section, we present an alternative method where the teleportation of these tripartite states is achieved using two Bell pairs as entanglement channel. In our scheme, we prepare two Bell pairs in the state $\left|\phi^{+}\right\rangle_{47}= \frac{1}{\sqrt{2}}(\left|00\right\rangle+\left|11\right\rangle)_{47}$
and $\left|\phi^{+}\right\rangle_{56}=\frac{1}{\sqrt{2}}(\left|00\right\rangle+\left|11\right\rangle)_{56}$
as entanglement channels. Similar to the previous scheme, Alice performs a Hadamard operation on qubit
$1$ of the state $ \ket{\Psi_B}_{123}$ and performs a single particle measurement on the same qubit. For the case where qubit $1$ collapse to the $\left|1\right\rangle_1$, the state of qubits (2,3) is given by $\left|{\Phi}\right\rangle_{23}= (a\left|10\right\rangle-b\left|00\right\rangle+c\left|01\right\rangle-d\left|11\right\rangle)_{23}$
and the total wave function of the system becomes $\left|{\Psi}\right\rangle_{tot}= \left|{1}\right\rangle_1 \otimes \left|{\Phi}\right\rangle_{23} \otimes \left|{\Psi}\right\rangle_{4567}$.

\begin{table}[ht]
\caption{Outcome of Alice's measurement and state of qubits $(6,7)$}
\label{tab:4}
\begin{tabular}{l l l l l }
\hline
State of & State of qubits (6,7) & State of qubits (6,7) \\
(2,5)\& (3,4) & when qubit 1 is in $\left|0\right\rangle_1$  & when qubit 1 is in $\left|1\right\rangle_1$ \\
\hline
$\left|{\phi^{+}}\right\rangle$  $\left|{\phi^{+}}\right\rangle$&$+a\left|10\right\rangle+b\left|00\right\rangle+c\left|01\right\rangle+d\left|11\right\rangle$&$+a\left|10\right\rangle-b\left|00\right\rangle+c\left|01\right\rangle-d\left|11\right\rangle$\\
 $\left|{\phi^{-}}\right\rangle$  $\left|{\phi^{+}}\right\rangle$ &$-a\left|10\right\rangle+b\left|00\right\rangle+c\left|01\right\rangle-d\left|11\right\rangle$&$-a\left|10\right\rangle-b\left|00\right\rangle+c\left|01\right\rangle+d\left|11\right\rangle$ \\

$\left|{\phi^{+}}\right\rangle$ $\left|{\phi^{-}}\right\rangle$  &$+a\left|10\right\rangle+b\left|00\right\rangle-c\left|01\right\rangle-d\left|11\right\rangle$&$+a\left|10\right\rangle-b\left|00\right\rangle-c\left|01\right\rangle+d\left|11\right\rangle$ \\
$\left|{\phi^{-}}\right\rangle$ $\left|{\phi^{-}}\right\rangle$&$-a\left|10\right\rangle+b\left|00\right\rangle-c\left|01\right\rangle+d\left|11\right\rangle$&$-a\left|10\right\rangle-b\left|00\right\rangle-c\left|01\right\rangle-d\left|11\right\rangle$\\

$\left|{\psi^{+}}\right\rangle$  $\left|{\phi^{-}}\right\rangle$&$+a\left|00\right\rangle+b\left|10\right\rangle-c\left|11\right\rangle-d\left|01\right\rangle$&$+a\left|00\right\rangle-b\left|10\right\rangle-c\left|11\right\rangle+d\left|01\right\rangle$\\
 $\left|{\psi^{-}}\right\rangle$ $\left|{\phi^{-}}\right\rangle$  &$-a\left|00\right\rangle+b\left|10\right\rangle-c\left|11\right\rangle+d\left|01\right\rangle$&$-a\left|00\right\rangle-b\left|10\right\rangle-c\left|11\right\rangle-d\left|01\right\rangle$\\

$\left|{\psi^{+}}\right\rangle$  $\left|{\phi^{+}}\right\rangle$&$+a\left|00\right\rangle+b\left|10\right\rangle+c\left|11\right\rangle+d\left|01\right\rangle$&$+a\left|00\right\rangle-b\left|10\right\rangle+c\left|11\right\rangle-d\left|01\right\rangle$\\
$\left|{\psi^{-}}\right\rangle$  $\left|{\phi^{+}}\right\rangle$&$-a\left|00\right\rangle+b\left|10\right\rangle+c\left|11\right\rangle-d\left|01\right\rangle$&$-a\left|00\right\rangle-b\left|10\right\rangle+c\left|11\right\rangle+d\left|01\right\rangle$\\

$\left|{\phi^{-}}\right\rangle$ $\left|{\psi^{+}}\right\rangle$ &$-a\left|11\right\rangle+b\left|01\right\rangle+c\left|00\right\rangle-d\left|10\right\rangle$&$-a\left|11\right\rangle-b\left|01\right\rangle+c\left|00\right\rangle+d\left|10\right\rangle$\\
$\left|{\phi^{+}}\right\rangle$ $\left|{\psi^{+}}\right\rangle$&$+a\left|11\right\rangle+b\left|01\right\rangle+c\left|00\right\rangle+d\left|10\right\rangle$&$+a\left|11\right\rangle-b\left|01\right\rangle+c\left|00\right\rangle-d\left|10\right\rangle$\\

$\left|{\phi^{-}}\right\rangle$  $\left|{\psi^{-}}\right\rangle$ &$-a\left|11\right\rangle+b\left|01\right\rangle-c\left|00\right\rangle+d\left|10\right\rangle$&$-a\left|11\right\rangle-b\left|01\right\rangle-c\left|00\right\rangle-d\left|10\right\rangle$\\
$\left|{\phi^{+}}\right\rangle$ $\left|{\psi^{-}}\right\rangle$ &$+a\left|11\right\rangle+b\left|01\right\rangle-c\left|00\right\rangle-d\left|10\right\rangle$&$+a\left|11\right\rangle-b\left|01\right\rangle-c\left|00\right\rangle+d\left|10\right\rangle$\\

$\left|{\psi^{-}}\right\rangle$ $\left|{\psi^{-}}\right\rangle$ &$-a\left|01\right\rangle+b\left|11\right\rangle-c\left|10\right\rangle+d\left|00\right\rangle$&$-a\left|01\right\rangle-b\left|11\right\rangle-c\left|10\right\rangle-d\left|00\right\rangle$\\
$\left|{\psi^{+}}\right\rangle$  $\left|{\psi^{-}}\right\rangle$&$+a\left|01\right\rangle+b\left|11\right\rangle-c\left|10\right\rangle-d\left|00\right\rangle$&$+a\left|01\right\rangle-b\left|11\right\rangle-c\left|10\right\rangle+d\left|00\right\rangle$\\

$\left|{\psi^{-}}\right\rangle$ $\left|{\psi^{+}}\right\rangle$&$-a\left|01\right\rangle+b\left|11\right\rangle+c\left|10\right\rangle-d\left|00\right\rangle$&$-a\left|01\right\rangle-b\left|11\right\rangle+c\left|10\right\rangle+d\left|00\right\rangle$  \\
$\left|{\psi^{+}}\right\rangle$  $\left|{\psi^{+}}\right\rangle$&$+a\left|01\right\rangle+b\left|11\right\rangle+c\left|10\right\rangle+d\left|00\right\rangle$&$+a\left|01\right\rangle-b\left|11\right\rangle+c\left|10\right\rangle-d\left|00\right\rangle$\\
\noalign{\smallskip}\hline
\end{tabular}
 \end{table}

 \begin{table}[ht]
\caption{Result of Alice's Bell measurement (BM) and corresponding unitary operations (UO) to be performed by Bob on qubits $(6,7)$}
\label{tab:3}
\begin{tabular}{l l}
\hline\noalign{\smallskip}
 BM& UO\\
result on & on $(6,7)$ \\
$(2,5)(3,4)$& after BM\\
\noalign{\smallskip}\hline\noalign{\smallskip}
$\left|\phi^{+}\right\rangle_{25}$ $\left|\phi^{+}\right\rangle_{34}$&$ I_6 \otimes I_7$\\
 $\left|\phi^{-}\right\rangle_{25}$ $\left|\phi^{+}\right\rangle_{34}$&$\sigma_{z_{6}} \otimes I_7 $\\

 $\left|\phi^{+}\right\rangle_{25}$ $\left|\phi^{-}\right\rangle_{34}$&$ I_6 \otimes \sigma_{z_{7}} $\\
 $\left|\phi^{-}\right\rangle_{25}$ $\left|\phi^{-}\right\rangle_{34}$& $\sigma_{z_{6}} \otimes \sigma_{z_{7}}$\\

 $\left|\psi^{+}\right\rangle_{25}$ $\left|\phi^{-}\right\rangle_{34}$& $\sigma_{x_{6}} \otimes \sigma_{z_{7}}$\\
 $\left|\psi^{-}\right\rangle_{25}$ $\left|\phi^{-}\right\rangle_{34}$&$i\sigma_{y_{6}} \otimes \sigma_{z_{7}} $\\

 $\left|\psi^{+}\right\rangle_{25}$ $\left|\phi^{+}\right\rangle_{34}$& $\sigma_{x_{6}} \otimes I_7$\\
$\left|\psi^{-}\right\rangle_{25}$ $\left|\phi^{+}\right\rangle_{34}$&$i\sigma_{y_{6}} \otimes I$\\

 $\left|\phi^{-}\right\rangle_{25}$ $\left|\psi^{+}\right\rangle_{34}$& $\sigma_{z_{6}} \otimes \sigma_{x_{7}} $\\
 $\left|\phi^{+}\right\rangle_{25}$ $\left|\psi^{+}\right\rangle_{34}$&$ I_6 \otimes \sigma_{x_{7}}$\\

 $\left|\phi^{-}\right\rangle_{25}$ $\left|\psi^{-}\right\rangle_{34}$&$\sigma_{z_{6}} \otimes i\sigma_{y_{7}}  $\\
$\left|\phi^{+}\right\rangle_{25}$ $\left|\psi^{-}\right\rangle_{34}$&$ I_6 \otimes i\sigma_{y_{7}} $\\

 $\left|\psi^{-}\right\rangle_{25}$ $\left|\psi^{-}\right\rangle_{34}$&$i\sigma_{y_{6}} \otimes i\sigma_{y_{7}}$\\
 $\left|\psi^{+}\right\rangle_{25}$ $\left|\psi^{-}\right\rangle_{34}$&$\sigma_{x_{6}}\otimes i\sigma_{y_{7}}$\\

 $\left|\psi^{-}\right\rangle_{25}$ $\left|\psi^{+}\right\rangle_{34}$&$i\sigma_{y_{6}} \otimes \sigma_{x_{7}}$\\
 $\left|\psi^{+}\right\rangle_{25}$ $\left|\psi^{+}\right\rangle_{34}$&$\sigma_{x_{6}} \otimes \sigma_{x_{7}}$\\
\noalign{\smallskip}\hline
\end{tabular}
\end{table}

\begin{table}
\caption{Comparision of Quantum Cost (QC), Gate count (GC) and classical bits (CB) used for the teleportation of tripartite state $\ket{\Psi_C}$. PP- proposed protocol}
\label{tab:5}
\begin{tabular}{l l l l l }
\hline
Scheme & Entanglement &QC & GC & CB\\
&channel &  &  & \\
\hline
Liu and Zhou\cite{zhong} & Five-qubit cluster state & $15$ & $19$& $4$\\
Chaudhury and Dhara\cite{BA16} & Four-qubit cluster state (3)& $ 20$& $32$& $6$\\
Wei, Zha and Yu\cite{WZY} & Four-qubit cluster state & $13$ & $17$ & $4$\\
Cai and Jiang\cite{CJ} & Four-qubit cluster state & $12$ &$ 16$ &$4$ \\
Our work $(PP_1)$ & Four-qubit cluster state & $12$ &$17$& $5$\\
Our work $(PP_2)$ & Two EPR pairs & $10$ &$15$& $5$\\
\hline
\end{tabular}
\end{table}
Now, Alice performs Bell measurements on particle pairs $(2,5)$ and $(3,4)$.
The results of Alice's Bell measurement and the state of qubits $(6,7)$ are shown
in table \ref{tab:4}. Depending on Alice's Bell measurement results, Bob performs unitary
operation on qubits (6,7) in his possession according to table \ref{tab:3}. The
quantum circuit used to implement this scheme is given in figure \ref{fig:3}. Finally, to recover the teleported state
$\ket{\Psi_B}$( $\ket{\Psi_D}$), Bob adds an ancilla qubit in the state $\left|1\right\rangle_8$ ($\left|0\right\rangle_8$)
 and performs two CNOT gates and a Z operation (based on Alice's information) on qubit 8 as shown in figure \ref{fig:3}. The receiver recovers the state in his qubits $(6,7,8)$ as $\ket{\Psi_B}_{867}$. Similarly, the teleportation of states $\ket{\Psi_A}$ and $\ket{\Psi_C}$ can be achieved with the same quantum circuit given in figure:\ref{fig:3} by entangling ancilla qubit prepared in the states $\ket{1}_8$ and $\ket{0}_8$ respectively. Note that, for the teleportation of these states, the CNOT gate between qubits 7 and 8 given in the dotted box in figure:\ref{fig:3} is removed.

\section{ Efficiency  Analysis and Comparison with other teleportation schemes}
\label{sec:3}
For a proper comparison with other schemes we need a common measure that quantifies all the resources used in the
teleportation scheme. A quantity known as quantum cost(QC) proposed by Perkowski \textit{et.al}\cite{perkowski}
quantifies the  entanglement resources, classical communication bits and unitary operations utilized in the scheme. It can be easily calculated by drawing an equivalent measurement-less circuit of the given scheme by removing the
measurement operators and replacing the double lines representing classical information in the circuit by single lines.
According to \cite{perkowski}, in a measurement-less circuit the quantum cost of all single qubit gates is zero, when
it is preceded or succeeded by a two-qubit gate and it is one otherwise. All two-qubit gates are assigned a quantum
cost of one. Gate count(GC) can also be used as a measure to quantify the resources consumed in a scheme, but the quantum
cost is considered to be a standard measure. Following this convention, one can find the quantum cost for
teleporting states $(\ket{\Psi_B})$ or $(\ket{\Psi_D})$ using four-qubit cluster state and by four-particle joint measurement proposed in section:\ref{sec:1} to be 13, the gate count 18 and classical communication bit (CB) to be 5 as shown in figure:\ref{fig:2}. Similarly, the quantum cost, gate count and classical bits required for teleporting states $(\ket{\Psi_A})$ or $(\ket{\Psi_C})$ through this scheme is 12, 17 and 5 respectively. Using the scheme proposed in section:\ref{sec:2}, where two Bell pairs and two Bell measurements are used, the QC, GC and CB for teleporting $(\ket{\Psi_B})$ or $(\ket{\Psi_D})$ are found to be 11, 16 and 5 respectively. For the states $(\ket{\Psi_A})$ or $(\ket{\Psi_C})$ these are 10, 15 and 5.

 By drawing the measurement-less quantum circuit representation of the related schemes presented in\cite{zhong}\cite{BA16}
 \cite{WZY} and  \cite{CJ}(where the teleported state is $\ket{\Psi_C}$)we have evaluated the quantum cost, gate count
 and classical bits used in the schemes and are tabulated along with the results in our proposed schemes in table:\ref{tab:5}. In \cite{zhong}, a five qubit entanglement  channel and five-particle joint measurement is used, where as  in \cite{BA16} three sets of four-qubit cluster states has  been used to complete the same quantum task. The quantum cost of these schemes shows that they are not as efficient as the  other schemes. The scheme \cite{WZY} presented  as a comment on
 \cite{BA16} shows better QC and GC values. In \cite{WZY} a CNOT gate applied between the four-qubit channel and
 an ancilla qubit makes it an effective five-qubit quantum channel. The scheme proposed in \cite{CJ} uses
 four-qubit cluster state as entanglement channel and effectively reduced the quantum
 cost by using two Bell measurements to complete the task. In our first method (PP1 in the table) we used four-particle joint measurement to do the same. But,  our second method (PP2)  demonstrates  that the QC for this task can be further reduced by two units when two Bell pairs  and two Bell measurements are used. These results  clearly indicate that our teleportation schemes have higher efficiency  and our second method completes the same quantum task with the minimal quantum cost.

\section{Conclusions}
We have presented two schemes for the teleportation of a class of three-particle states.The first method uses
four-qubit cluster state and the second method employs two Bell pairs as entanglement channels. The quantum
circuits for the proposed schemes are given explicitly in
figures: \ref{fig:2} and \ref{fig:3}. Efficiency analysis comparison of the quantum cost, gate count and
classical bits of our scheme with
other related schemes in table: \ref{tab:5} shows that both of our schemes are more efficient compared with the other schemes.
We have found that by employing two Bell pairs and Bell measurements one can achieve the same quantum task
of teleporting tripartite states discussed in section:\ref{sec:1} with lesser quantum resource. Our second scheme adopts
a practically feasible method for teleporting these states using Bell pairs and Bell measurements.

\section{Acknowledgements}
We thank Kerala State Council for Science, Technology and Environment for providing  financial support.

\end{document}